\documentclass{ws-procs961x669}            
\usepackage{hyperref}

\def\be{\begin{eqnarray}}
\def\ee{\end{eqnarray}}

\begin{document}
\title{Dark matter properties
from the Fornax globular cluster timing:\\ dynamical friction  and cored profiles}

\author{D. Blas$^*$}

\address{{Grup  de  F\'isica  Te\`orica,  Departament  de  F\'isica,\\ Universitat  Aut\`onoma  de  Barcelona,  08193  Bellaterra, Spain}, and \\
{Institut de Fisica d'Altes Energies (IFAE),\\ The Barcelona Institute of Science and Technology,
Campus UAB, 08193 Bellaterra , Spain}\\
$^*$E-mail: \href{dblas@ifae.es}{dblas@ifae.es}}

\begin{abstract}
I summarize  our recent results to use the orbits of globular clusters (GCs) in the Fornax dwarf spheroidal (dSph) galaxy to learn more about dark matter (DM) properties.  Our focus is on clarifying how dynamical friction (DF) from the DM halo is modified from the different microscopic properties of DM, which may alter {\it both} the scattering processes responsible of DF and the DM profiles (in particular generating a core), which also modifies DF.  
We consider: $(i)$ fermionic degenerate dark matter (DDM), where Pauli blocking should be taken into account in the dynamical friction computation;  $(ii)$ self-interacting dark matter (SIDM) and $(iii)$  ultralight dark matter (ULDM), for which this problem has been addressed by a variety of methods in recent literature.  
We derive DF with a Fokker-Planck formalism, reproducing previous results for ULDM and cold DM, while providing new results for DDM. 
Furthermore, ULDM, DDM and SIDM   may generate cores in dSphs, which  suppress  dynamical friction and prolong GC orbits. 
We conclude that in all these cases the modifications in the DM modelling does not easily solve the  so-called timing `problem' of Fornax GCs. 
We finally study this `problem' in terms of the initial conditions, demonstrating that the observed orbits of Fornax GCs are consistent with this expectation of a cuspy DM profile with a mild  `fine-tuning' at the level of $\sim25\%$. 
\end{abstract}


\bodymatter

\section{Introduction and motivation}\label{aba:sec1}

The presence of dark matter (DM) in the universe is probed by several different methods. In particular, the rotation curves of \emph{tracers} of the gravitational potential in galaxies has long been considered as one of the most pressing motivations for the existence of a halo of  DM  extending beyond the volume occupied by visible matter. When tracers move in this \emph{medium}, their dynamics may also be altered by direct momemtum exchange with the DM, which may lead to further dynamical consequences beyond rotation curves, as for instance tidal disruptions, dynamical heating or dynamical friction. The latter is the focus of this contribution, and in particular using it to learn about properties of DM.

Dynamical friction is generated by the relative motion between  the tracer (probe) and the DM medium.  Indeed, as the probe moves in the DM halo, its gravitational interaction with the latter leaves behind an overdensity of DM particles (wake) that pulls back gravitationally from it. The net result is a friction force, first calculated by Chandrasekhar~\cite{Chandra43}, and that can be parameterized as (for a probe of mass $m_\star$)
\cite{BinneyTremaine2}
\be
\frac{d\mathbf{V}}{dt} &=&-\frac{4\pi G^2 m_\star\rho}{V^3}\,C\,{\bf V}, \label{eq:DF}
\ee
where $\mathbf{V}$ is the relative velocity, $G$ is Newton's constant, $\rho$ is the energy density of the DM medium and $C$ is a parameter depending on the way momentum is exchanged between the probe and the DM medium. For a classical gas of particles with mass $m$, and with a cold distribution in velocities $v_m$ given by $f(v_m)$, the later reads
\be\label{eq:Chand} C_{\rm class}&=&4\pi\ln\Lambda\int_0^Vdv_mv_m^2f_v(v_m).\ee
In this formula, $\ln \Lambda$ represents the so-called Coulomb logarithm, and is a factor of $O(1)$. We can already understand that different models of DM may alter the previous formula through modifications of the microscopic scattering (for instance if DM is not a simple collection of free classical particles, but has some coherent or quantum properties), different DM energy densities $\rho$ or  distribution functions $f_v$. 

The question we want to address is if these features can lead to observable consequences. For this, let us consider a system where DF is supposed to have played a relevant role in current observations: the globular clusters (GCs) of the Milky Way dwarf spheroidal (dSph) satellite galaxies. The latter are 
believed to be DM dominated `compact' galaxies \cite{walker2009universal,Cole2012}. One intriguing puzzle about the dSph galaxies related to DF concerns the GCs of the Fornax dSph \cite{Tremaine1976a}.  Fornax is a very luminous nearby (kpc away) dwarf satellite, with a stellar mass of around $\sim 4\times  10^7\, M_{\odot}$ and of~$\sim$ kpc scale. It contains six known GCs \cite{Cole2012,wang2019rediscovery}, with masses around $m_\star \sim 10^5\, M_{\odot}$, which we consider as our `probes' moving in the DM halo. The puzzle arises because  2 of the most innermost GCs  should have lost enough momentum  due to DF to make them fall to the center of Fornax, while they live relatively far away from this point. In other words, from the expression \eqref{eq:DF}, one can na\"ively estimate the typical time scale to fall to center for GCs, 
 \be
 \tau \equiv \frac{|\mathbf{V}|}{|d \mathbf{V} / d t|} \sim 1.8\left(\frac{V}{12 \mathrm{~km} / \mathrm{s}}\right)^{3}\left(\frac{10^5\, M_\odot}{m_\star}\right)\frac{2 \times 10^{7} \frac{M_{\odot}}{\mathrm{kpc}^{3}}}{\rho}\ \mathrm{Gpc}. \label{eq:tau}
 \ee
 When applied to the six GCs of Fornax  assuming the usual CDM cusp density profile (see, e.g. \cite{Meadows20}),  this time scale for the six GCs is presented in Table~\ref{aba:tbl1}\cite{Bar:2021jff} (see also  \cite{Hui2017}). 
On the other hand, the stellar content of the GCs (and most of the stellar content of Fornax) is old, with life estimates $ >10 ~$Gyr \cite{de2016four,Mackey2003a} \cite{del2013spatial,wang2019morphology}. Hence these GCs had enough time to fall to the center since their `birth'.

 \begin{table}
\tbl{Some details of Fornax GCs: mass, projected radius and CDM instantaneous DF time (Eq.~\eqref{eq:tau}). See Ref.~\cite{Bar:2021jff} for more details about how these numbers are obtained and the origin of the discrepancies with previous literature.}
{\begin{tabular}{@{}c|ccc@{}}
\toprule
& $m_{\star}\left[10^{5} M_{\odot}\right] $& $ r_{\perp}[\mathrm{kpc}] $& $\tau_{\mathrm{CDM}}[\mathrm{Gyr}] $\\
\hline \text { GC1 } &$ 0.42 \pm 0.10 $& $1.73 \pm 0.05 $& 119 \\
\text { GC2 } & $1.54 \pm 0.28 $& $0.98 \pm 0.03 $& 14.7 \\
\text { GC3 } & $4.98 \pm 0.84$ & $0.64 \pm 0.02 $&  2.63 \\
\text { GC4 }  & $0.76 \pm 0.15 $& $0.154 \pm 0.014 $& 0.91 \\
\text { GC5 }  & $1.86 \pm 0.24 $& $1.68 \pm 0.05$ & 32.2 \\
\text { GC6 } & $\sim 0.29$ & $0.254 \pm 0.015$ & 5.45
\end{tabular}
}
\label{aba:tbl1}
\end{table}

 This puzzle (in particular the time scales for GC3 and GC4) has been highlighted as a possible tension of the standard DM paradigm that may be solved by changing the properties of DM, see e.g. \cite{Hui2017}. A first important observation is that  the Jeans analysis based on kinematic information of Fornax (the velocity dispersion along the line of sight of $\sim 2500$ stars as a function of radius) is compatible with the energy density profiles M19 NFW and M19 ISO  expected from standard CDM \cite{Meadows20}, but also with the presence of a cored distribution arising in different interesting DM models. The latter include  fermionic DM of masses where quantum degeneracy may be relevant (DDM) in the dynamics of Fornax \cite{Domcke2015,Randall2017,DiPaolo2018,Savchenko2019,Boyarsky2009,Alvey:2020xsk}, ultra-light DM  with de Broglie wavelength large enough to affect the dynamics of Fornax\cite{Hui2017} or self-interacting DM (SIDM) within the parameter space allowed by other observables, e.g. \cite{Kaplinghat:2015aga,Tulin:2017ara}. A key motivation for our work is exploring whether DF may distinguish among these options, and hint towards new properties of DM.

\section{Derivation of DF for different DM models with a Fokker-Planck approach}\label{aba:sec2}

In order to include non-trivial properties of the DM microphysics into the expression \eqref{eq:DF}, we followed a  Fokker-Planck (FP) approach in  \cite{Bar:2021jff}, where the probe particle (a GC in our case)  corresponds to species $1$ traveling through a gas of spectator particles (species $2$) with a certain distribution  $f_2$ (DM particles in our case). 
We consider the following elastic scattering process of two particle species, 
$$1 (p) + 2 (k) \to 1(p') + 2(k').$$
The phase space distribution function for the particle species $1$ (the GC) evolves according to the Boltzmann equation,
\be \frac{df_1}{dt} = C[f_1].\ee
The collision integral $C[f_1]$ contains information about the elastic scattering processes, and is written as
\be
&&C[f_1] = \frac{(2\pi)^4}{2E_p} \int d\Pi_{k} d\Pi_{p'} d\Pi_{k'} \, 
 \delta^{(4)}(p+k - p' - k') |\overline{{\cal M}}|^2 
\nonumber \\
&&
\times \Big[ 
f_1(p') f_2(k') (1 \pm f_1(p) ) ( 1  \pm f_2(k) ) - f_1(p) f_2(k) (1 \pm f_1(p') ) (1 \pm f_2(k') )
\Big],
\ee
where $|\overline{\cal M}|^2$ is a squared matrix element averaged over initial and final spins, and $d\Pi_k = \frac{g}{2E_k} \frac{d^3k}{(2\pi)^3}$ is the Lorentz invariant phase element with the number of internal degrees of freedom $g$. The sign in $1\pm f_i$ refers to bosons ($+$) or fermions ($-$), respectively. The Boltzmann equation can be greatly simplified if the momentum exchange
\be q = p' - p,
\ee
 is smaller than the typical momentum in the distribution function $f_1$.  This is the adequate limit to compute DF \cite{BinneyTremaine2}.
In such cases, the Boltzmann equation reduces to the nonlinear Fokker-Planck equation,
\be
\frac{df_1}{dt} &=&
- \frac{\partial}{\partial p^i} \left[ f_1 (1\pm f_1) D_i \right]+\frac{1}{2} \frac{\partial}{\partial p^i} \left[ \frac{\partial }{\partial p^j} (D_{ij} f_1) \pm f_1^2 \frac{\partial}{\partial p^j} D_{ij} \right],\label{eq:FP}
\ee
where the diffusion coefficients are defined in \cite{Bar:2021jff,BinneyTremaine2}.
Furthermore, the gravitational scattering of a probe particle of mass $m_\star$ and a particle in the medium with mass $m$ is described by the spin-averaged matrix element
\be
| \overline {\cal M}|^2 = \frac{1}{2 s + 1} \frac{(16\pi G)^2 m^4 m_\star^4}{\left[(q^{0})^2-{\bf q}^2\right]^2},
\label{me}
\ee
where $s$ is the spin of the particle in the medium, and $(q^0,\bf q)$ is the transferred $4-$momentum.
In the nonrelativistic limit, we can neglect $q^0$ and maintain only ${\bf q}$ in Eq.~(\ref{me}).

Of particular importance for our analysis is the diffusion coefficient $D_{||}$, corresponding to the diffusion in momentum parallel to the probe object's instantaneous velocity. Indeed, the DF deceleration on a probe of mass $m_\star$ moving with  velocity ${\bf V}$ w.r.t. the medium is
\cite{BinneyTremaine2}
\be
\frac{d\mathbf{V}}{dt} &=& \frac{D_{||}}{m_\star}\hat{\mathbf{V}}.
\ee
From \eqref{eq:DF}, we see that the dimensionless coefficient $C$  of the DF reads,
%
\be C&=& -\frac{V^2 D_{||}}{4\pi G^2m_\star^2\rho}.\ee 

When computed for a gas of classical particles (where $f_2$ corresponds to a classical distribution), one reproduces the known results of the  Chandrasekhar~\cite{Chandra43} formula, with $C$ given by \eqref{eq:Chand}. This limit is also the relevant one for the SIDM case of interest here, since the corresponding cross-sections always correspond to large mean-free paths, where the approximation of the  Chandrasekhar calculation holds.

Regarding DDM, when one assumes that $f_2$ is given by a Fermi-Dirac distribution close to the degeneracy limit\footnote{Note that it is not known what is the momentum distribution for degenerate fermions interacting only gravitationally. Still, one normally assumes an equilibrium configuration as a first approximation. This assumption may also be used to constrain the mass of fermionic DM from virialized DM objects \cite{Alvey:2020xsk,Tremaine:1979we}.} one finds 
\be\label{eq:Cdeg} C_{\rm DDM}&\to&\ln\Lambda\begin{cases}
	1 & V\gg v_F \\ \frac{V^3}{v_F^3} & V\ll v_F
\end{cases},\ee
where $v_F$ is the Fermi velocity, related to the DM energy density by 
\be\label{eq:rhoddm}\rho&=&\frac{g m_{\rm DM}^4v_F^3}{6\pi^2},\ee
where $g$ represents the number of degrees of freedom of the species (see also \cite{Chavanis2020landau}). We used $m_{\rm DM}$ for the DM mass. This modification of the DF formula changes the  falling time  $\tau$, though not parametrically.  Furthermore,  the DDM case also generates a \emph{core} due to the quantum degeneracy pressure. Both effects are useful to reduce DF and hence prolong $\tau$. 

A similar calculation can be done for the bosonic case with large occupation numbers. In this case,  the collision term is also modified and allows our formalism to capture the most relevant phenomena for DF in the ULDM case described in \cite{Hui2017,Lancaster2020,Bar-Or:2018pxz,Bar-Or:2020tys}. Indeed,  these  Bose-enhancement terms in \eqref{eq:FP} generate  large-scale density fluctuations, causing additional velocity drift that can be characterised by an extra term to $C\to C+\Delta C$ as%
\be
\Delta C &=& \ln \Lambda \left( \frac{m_{\rm eff}}{m_\star} \right) 
\bigg( {\rm erf}(X_{\rm eff}) - \frac{2 X_{\rm eff}}{\sqrt{\pi}} e^{-X_{\rm eff}^2} \bigg),
\ee
where $m_{\rm eff} = \pi^{3/2} \rho / (m_{\rm DM}\sigma)^3$ is the ULDM mass enclosed in an effective de Broglie volume and $X_{\rm eff} \equiv v/\sqrt{2} \sigma_{\rm eff}$ with $\sigma_{\rm eff} = \sigma/\sqrt{2}$. Numerically, $ m_{\rm eff}\approx 1.2\times 10^{6}\left(10^{-21}~{\rm eV}/m_{\rm DM}\right)^3[\rho/(3\times 10^{7}~M_{\odot}/{\rm kpc}^3)][(10~{\rm km/s})/\sigma]^3$~M$_\odot$. With these numbers and keeping in mind a typical GC mass $m_\star\sim10^5$~M$_\odot$, the $\Delta C$ effect becomes quantitatively important in Fornax for $ m\lesssim 3\times 10^{-20}$~eV. This value starts to be in tension with other constraints of ULDM, see e.g. \cite{Bar:2018acw,Marsh:2018zyw}. Notice that the previous calculation is correct in the limit $r>\lambda_{\rm db}$, where $\lambda_{\rm db}$ is the de Broglie wavelength of DM. In the opposite regime, one can have effects coming from the coherent nature of the DM waves which affect the scattering cross-section~\cite{Hui2017}, or even resonant processes that our formalism does not capture, see e.g.~\cite{Wang:2021udl,Bar-Or:2020tys}.

\section{Timing of Fornax GCs for different DM models}\label{aba:sec3}

\begin{figure}[h]
\hspace{-.4cm}\includegraphics[width=0.49\linewidth]{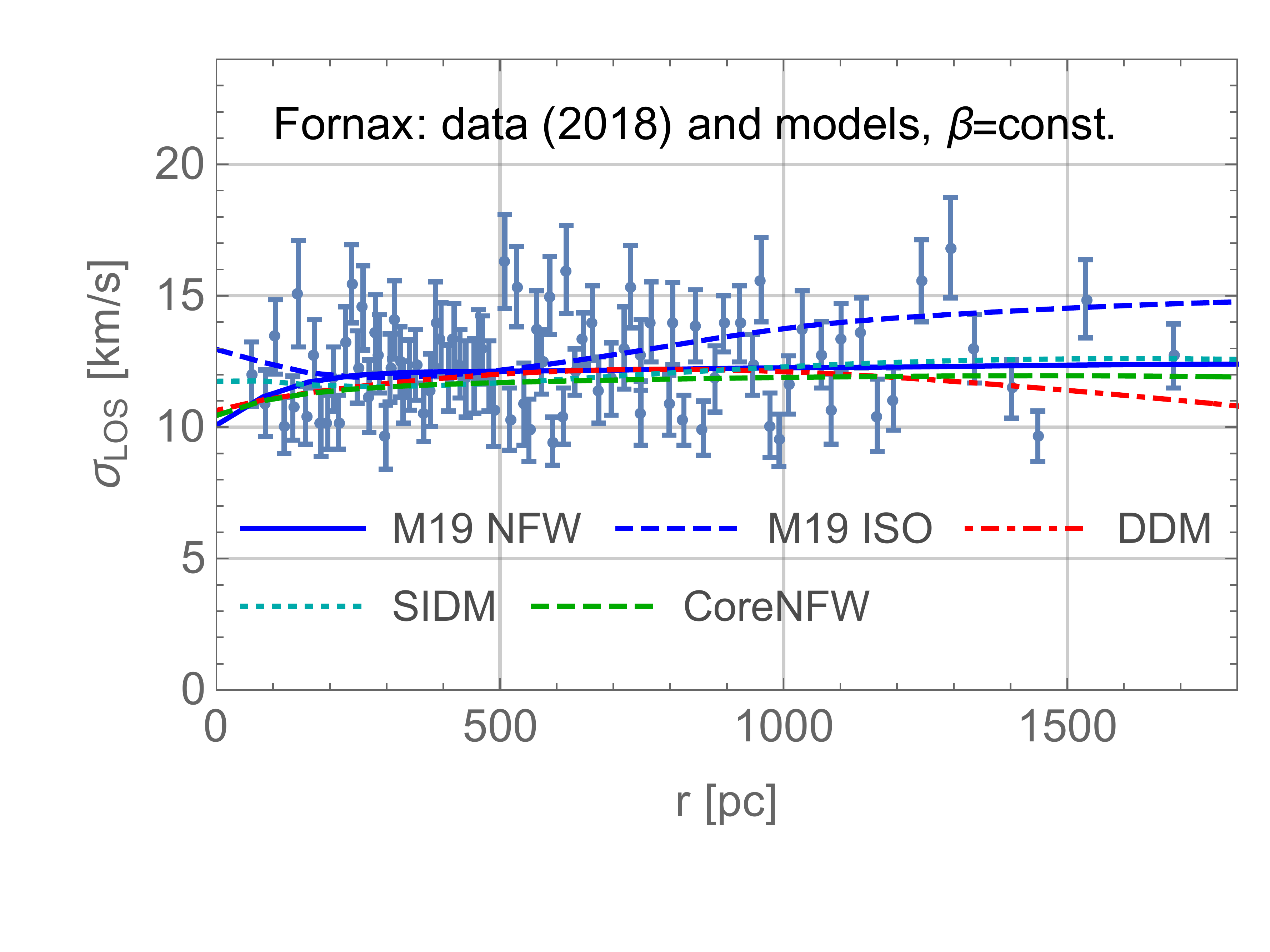}~~~~
  \includegraphics[width=0.49\linewidth]{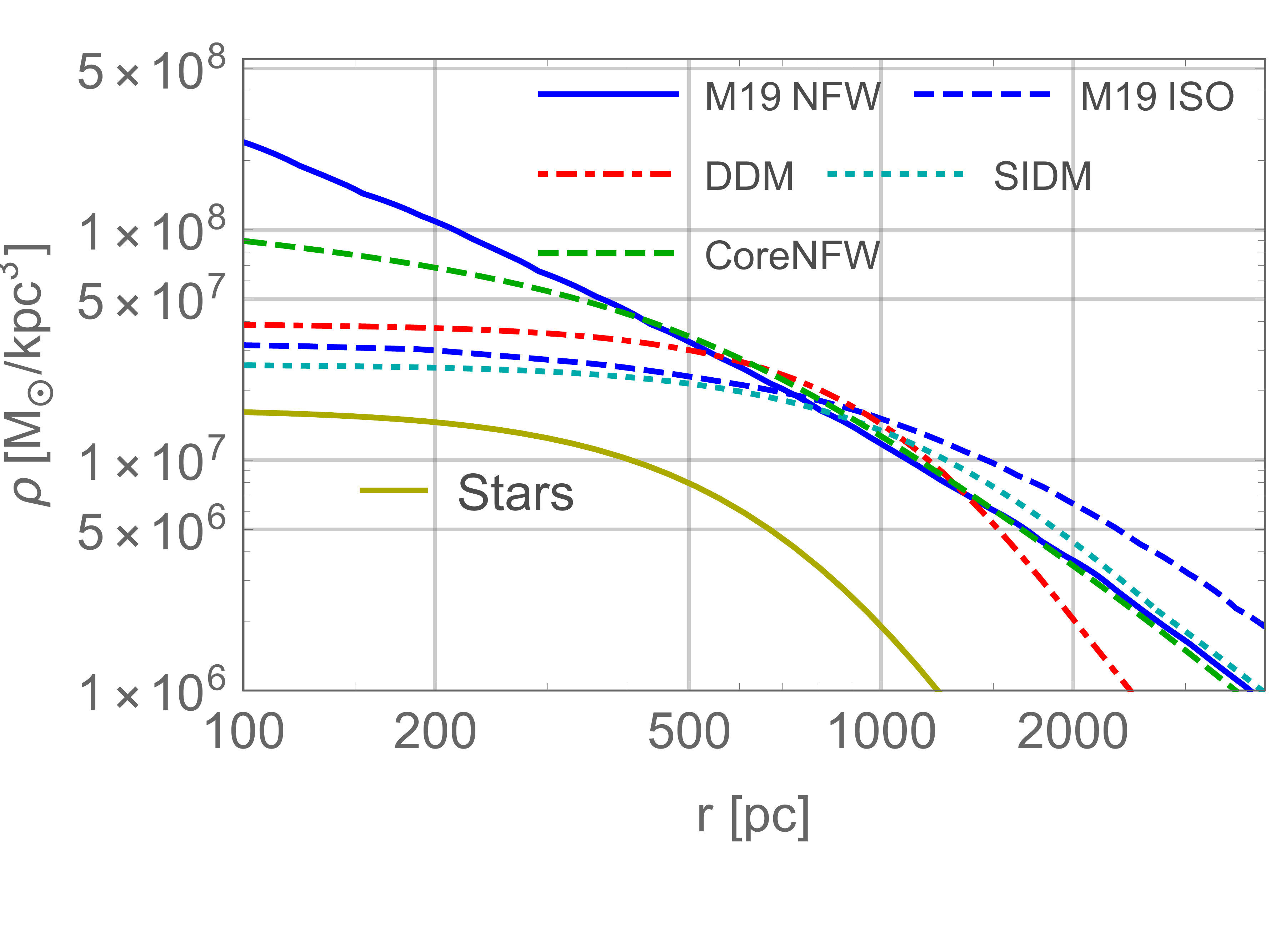}
\vspace{-.5cm}
\caption{{\bf Left panel}: LOSVD data and fits for the M19 NFW and M19 ISO of \cite{Meadows20},  DDM with mass $m\approx 135\,$eV and SIDM models with cores of order kpc. {\bf Right panel}: Density profiles corresponding to these cases. The energy density of the star content of Fornax is also shown.}
\label{fig:prof}
\end{figure}

From the previous calculation, one can estimate the falling time of the GCs in Fornax once the profile of DM energy density is known. For the latter, one can use a Jeans method consistent with the line-of-sight velocity (LOSV) observed for Fornax~\cite{BinneyTremaine2}. However, as already stressed, the latter is equally well fitted by a M19 NFW, M19 ISO CDM profiles, as long as DDM of mass $m_{\rm DM}\approx 135$\, eV (which should generate a degenerate core of $~$kpc) 
  or a SIDM model with velocity averaged cross section $\langle\sigma v\rangle / m_{\rm DM} \sim 25 \mathrm{~cm}^{2} \mathrm{~g}^{-1} \mathrm{~km} \mathrm{~s}^{-1}$  (also generating a  isothermal  cored profile of kpc size, this time due the
SIDM scatterings \cite{Kaplinghat:2015aga}). These cases are shown, together with the relevant data of the dispersion in the LOSV, $\sigma_{LOS}$, as a function of radius in the left panel of Fig.~\ref{fig:prof}. It should be clear from this panel that this data agrees well with the models just described. The right panel shows the different energy density profiles for these cases also as a function of radius. For completeness we have also shown the energy density of the star content of Fornax, to show explicitly that DM is needed in this system. Another CDM model (coreNFW) is also shown. These models aims at describing a DM dominated halo where the baryonic feedback is somehow taken into account.

\begin{figure}
\centering
\includegraphics[width=0.49\linewidth]{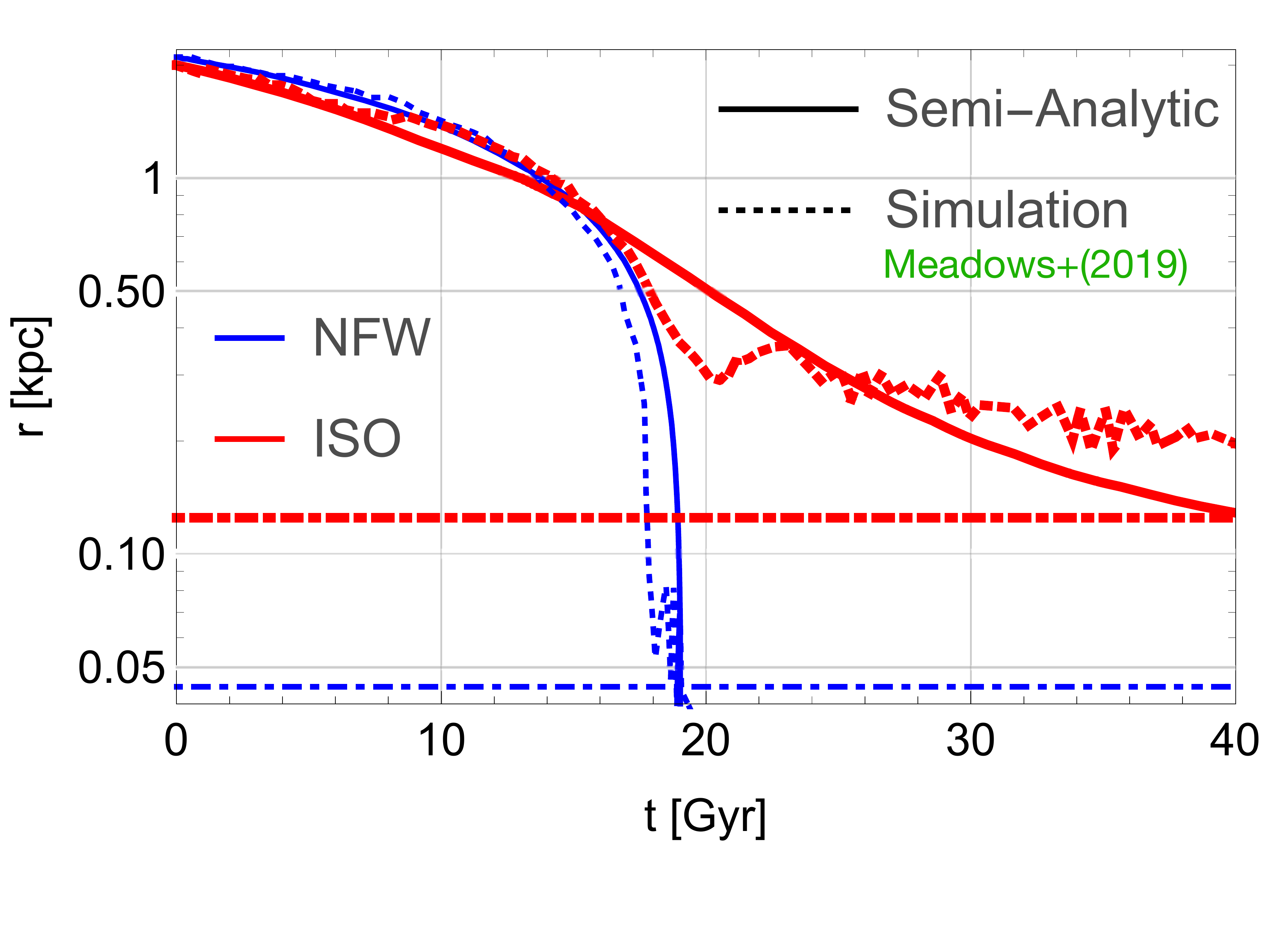}
\vspace{-.5cm}
\caption{Radius of an infalling orbit of a GC with mass $m_\star = 3\times10^5\, M_\odot$. In dotted blue (thick
dotted red) we plot the simulation result in \cite{Meadows20}
for the NFW (ISO) halo. In solid lines, we plot our semi-
analytic integration. Horizontal dot-dashed lines show the radii where  the semi-analytic
treatment breaks down.}
\label{fig:orbits}
\end{figure}

Now that we have all the ingredients to compute the DF, we proceed to find the orbital motion of the different GCs of Fornax by solving the equation
\begin{equation}
\frac{d \mathbf{V}}{d t}=-\frac{G M(r)}{r^{2}} \hat{r}-\frac{4\pi G^2 m_\star\rho}{V^3}\,C\,{\bf V}, \label{eq:dft}
\end{equation}
for the relevant cases. This is a `semi-analytical' approach that  yields good agreement with simulations \cite{Bar:2021jff}. The results are only mildly dependent on the Coulomb logarithm $\ln \Lambda$, which we calibrate to numerical work in \cite{Bar:2021jff}. As an example, we show in Fig.~\ref{fig:orbits} the evolution of the radius as a function of time, as compared to two profiles studied numerically in \cite{Meadows20}. 

\begin{figure}
\centering
\includegraphics[width=0.49\linewidth]{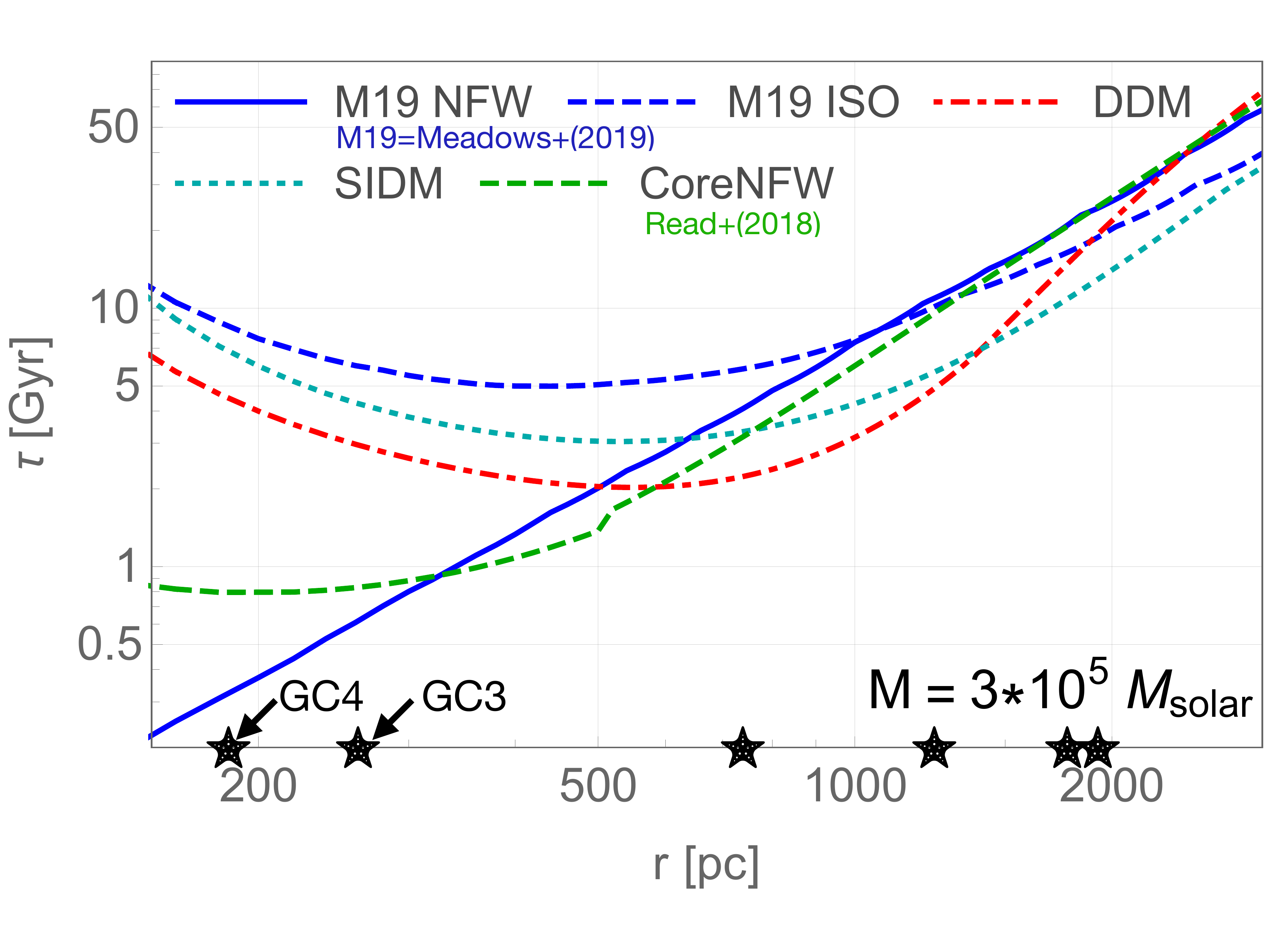}
\vspace{-.5cm}
\caption{ Instantaneous DF time, evaluated for different DM models for 
 $m_\star = 3 \times 10^5 M_\odot$ as a function of  distance to the center of Fornax. The stars represent the positions of the different GCs. }
\label{fig:times}
\end{figure}

From the previous analysis, we are now in the position to find the time $\tau$ that it takes to a GC to fall to the center of Fornax for different DM models. We will address the initial conditions momentarily. For now, let us compare the `instantaneous'  time in the DF formulae for the different GCs in CDM, DDM of mass $m_{\rm DM}\approx 135$\,eV and SIDM generating a kpc core. This is shown in Table~\ref{aba:tbl1} and Fig.~\ref{fig:times}. The conclusion is clear: both models beyond CDM prolong the plunging time of the problematic GCs, and hence alleviate the timing problem. 

\begin{table}
\tbl{Instantaneous DF times for the GCs of Fornax in the DDM and SIDM  models described in the main text.}
{\begin{tabular}{@{}c|ccc@{}}
\toprule
& $\tau_{\mathrm{CDM}}[\mathrm{Gyr}] $&$ \tau_{\text {DDM }}^{(135)}[\mathrm{Gyr}] $&$ \tau_{\text {SIDM }}[\mathrm{Gyr}]$\vspace{.1cm} \\
\hline \text { GC1 } &  119 & 122 & 79.3 \\
\text { GC2 } &  14.7 & 7.12 & 8.82 \\
\text { GC3 } &  2.63 & 1.48 & 2.21 \\
\text { GC4 } & 0.91 & 10.7 & 14.8 \\
\text { GC5 } &  32.2 & 30.1 & 20 \\
\text { GC6 } & 5.45 & 16.1 & 22\\
\end{tabular}
}
\label{aba:tbl1}
\end{table}

This conclusion is confirmed by a numerical study of the orbits, where the different DM profiles and dispersion properties are taking into account for the different models.  The latter analysis is vital to find the real dynamics of the GCs, since the timescales can vary by $O(1)$ factors. We will skip the details of this study in this short contribution, and the interested reader is invited to check our work \cite{Bar:2021jff} for details. Let us simply mention that beyond the numerical integration of Eq.~\eqref{eq:dft}, in \cite{Bar:2021jff} we also considered different projection effects that alter the radius of the different GCs and their velocities.  The main lesson we learned is that  the conclusions extracted from the estimates in Table~\ref{aba:tbl1} and Fig.~\ref{fig:times} are robust against these uncertainties.

The main lesson from Fig.~\ref{fig:times} is that cores (including those in coreNFW) increase the time $\tau$, but to make GC4's settling time substantially different, one requires a large (kpc-size) core. Note that the microscopic properties in the DF formula are not particularly relevant in our calculations. 

Before closing this section, let us notice that in Table~\ref{aba:tbl1} and Fig.~\ref{fig:times} we have focused on DM models which help prolonging the orbits of GCs in Fornax. However, for the DDM case this requires models with masses already in strong tension with Ly-$\alpha$ observations \cite{Baur:2015jsy} (and light tension with Jeans analysis of other systems \cite{Alvey:2020xsk}).  In fact, regarding the  Ly-$\alpha$ constraints, the interesting reader may also find in our work \cite{Bar:2021jff} a rather model independent bound from these observations 
\be
\label{eq:ddmcosmobound}
 gm_{\rm DM}^4&>&2\times\left(1.4~{\rm keV}\right)^4,
 \ee
confirmed in the recent detailed  work \cite{Carena:2021bqm}. On the contrary, the SIDM model we used for  the existence of a  core at kpc  distances can be read from the relation
\be
r_{c} \sim \frac{m_{\rm DM}}{\rho \sigma}=48 \frac{10^{8} M_{\odot} / \mathrm{kpc}^{3}}{\rho} \frac{1 \mathrm{~cm}^{2} / \mathrm{gr}}{\sigma / m_{\rm DM}} \mathrm{kpc}\,,
\ee
related to the isothermal profile generated by scattering of DM particles with cross-section $\sigma$. The ballpark used in the previous formulae corresponds to viable models of DM \cite{Kaplinghat:2015aga,Tulin:2017ara},  which puts our analysis on a solid phenomenological basis. Indeed, one is tempted to conclude that the Fornax GCs favour the SIDM model.

\section{Brief discussion on the late-time distribution of GCs}\label{sec:CDF}

\begin{figure}[h]
\hspace{-.4cm}\includegraphics[width=0.49\linewidth]{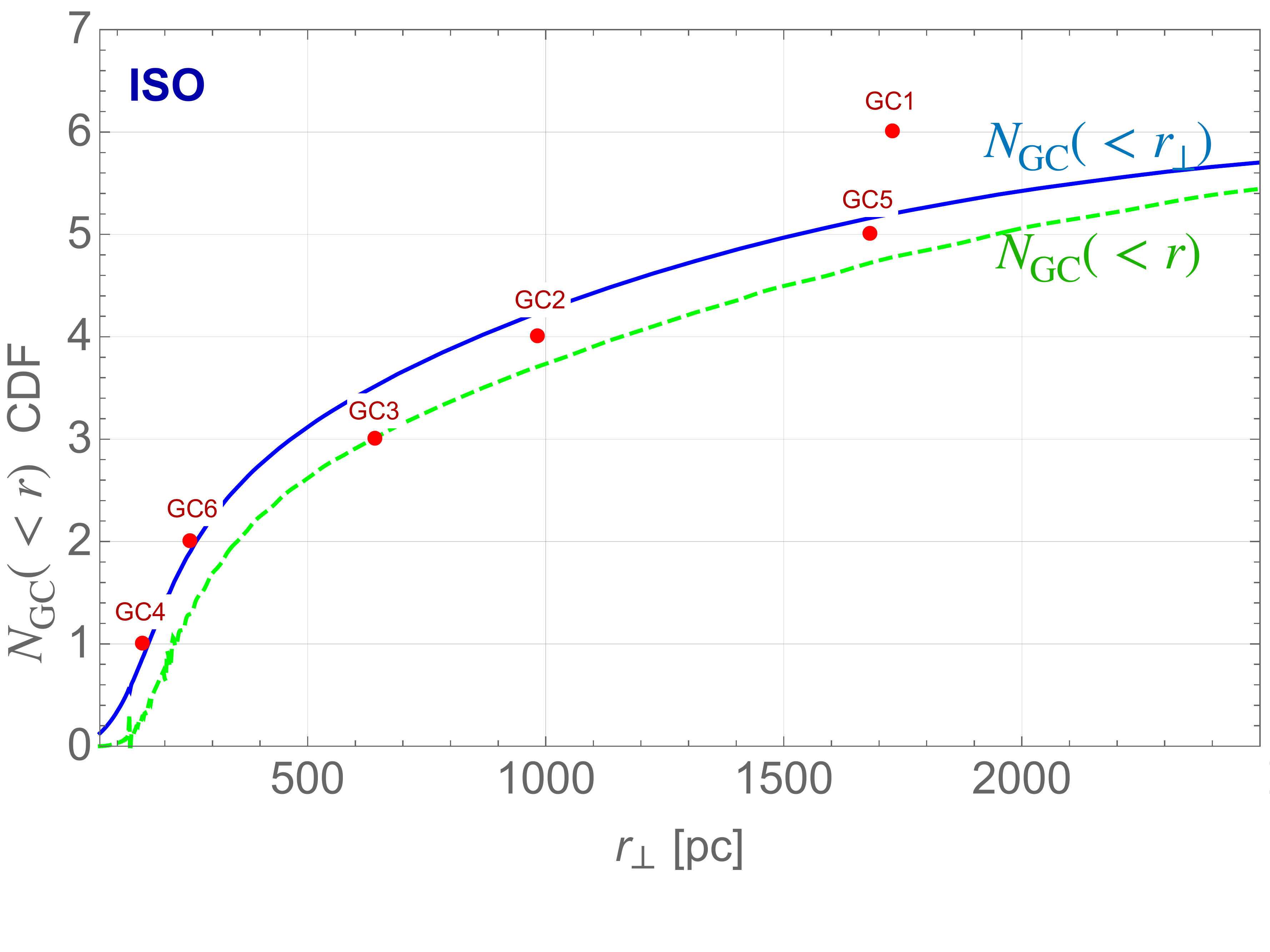}~~~~
  \includegraphics[width=0.49\linewidth]{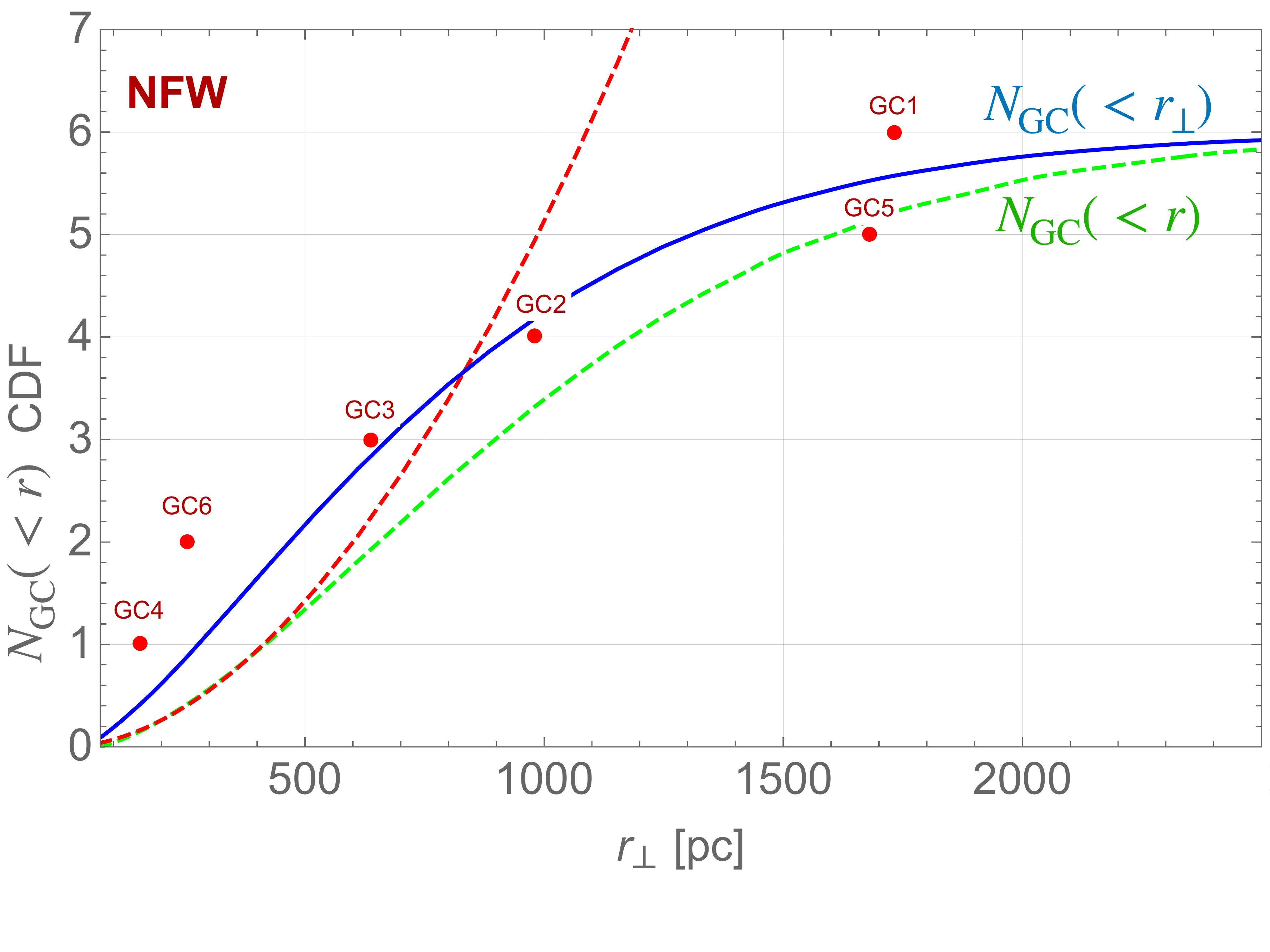}
\caption{CDF (number of clusters at $r_\perp$ below the value of the $x$ axes) of projected radii using $\tau(r)$ from Fig.~\ref{fig:times}. {\bf Left:} Cored ISO halo. {\bf Right:} NFW halo. 
The solid blue line shows the CDF after projection effects are taken into account. The dashed green line shows the result before
projection. For the NFW case, the small-$r$ prediction of Eq.~\eqref{eq:smallr} is shown by the red dashed line. Observed Fornax GCs are
also shown. The initial radial GC PDFs used to make the plot are explained in  \cite{Bar:2021jff}.}
\label{fig:CDF}
\end{figure}

Our discussion has so far focused on the falling time of the innermost GCs of Fornax. The `puzzle' we are trying to address relies on this time being smaller than what one would naively expect for arbitrary GCs that have lived in the dSph DM halo for long enough. However, given a collection of GCs, one expects some of them to be form currently at distances that  could seem fine tuned for the average GC. Hence, the best-posed question to learn about the DM effects in the dynamics of GCs in Fornax (given that we observe 6 of them) is which is  the long-term distribution function of these objects as a function of radius and for different initial conditions. Note first,  that the problem of initial conditions has already been identified as a candidate to explain the phenomenology we are discussing~\cite{Shao:2020tsl,Meadows20}. Here we will summaryze the \emph{analytical} treatment developed in \cite{Bar:2021jff}. For this, the key analytical tool is the time it takes an object to move from radius $r_0$ to $r_f$. In the case of nearly-circular orbits, this time can be estimated as
\be
\Delta t\left(r_{i} ; r_{f}\right)=\int_{r_{f}}^{r_{0}} \frac{d r}{2 r}\left(1+\frac{d \ln M}{d \ln r}\right) \tau\left(r, v_{\text {circ }}(r)\right).
\ee
Quite remarkably, one can show that for objects that today are at small radii $r\ll r_{\rm crit}$, with $r_{\rm crit}$ parametrizing the radius such that any object that \emph{started} its life at $r< r_{\rm crit}$ as already fallen into the center, the conditional distribution function (CDF) of GCs as a function of radius in a NFW profile satisfies 
\be
F_{\Delta t}(r) \approx A \frac{\tau(r)}{\Delta t}, \label{eq:smallr}
\ee
independently on initial conditions. This prediction is modified in the case of cored profiles. To have a better picture of this effect, we show in Fig.~\ref{fig:CDF} the number of GCs expected to be enclosed at projected radius $r_\perp$, when numerically integrating the orbital motion with our semianalytical model (no need of $N$-body simulations). The initial conditions we used for these plots are discussed in  \cite{Bar:2021jff}. They are not very particularly relevant for this discussion since we found that the main features of  Fig.~\ref{fig:CDF} are reproduced by all reasonable initial conditions we considered. 

A visual comparison of the left panel in Fig.~\ref{fig:CDF} (performed with an isothermal cored for the model of SIDM described above) and the right panel (performed with NFW) shows that GC4, GC6 and GC3 tend to align better over the CDF of the second one, in particular after projection effects are taken into account (blue-solid line). As we just mentioned, these figures were produced for one of many possible initial conditions that reproduce reasonably well the distribution of final GCs, and we could not find any other reasonable initial data which could account for the position of GC4 today without destroying the agreement of the rest of GCs.
Still, these figures also show that the level of `fine-tuning' to explain GC4, GC6 and GC3 within the standard NFW profile is not very problematic, which implies that a solid conclusion about the role of DF in the final distribution of GCs requires the observation of (several) other similar systems. Another interesting observation is that any reasonable distribution agreeing with current data should have included several inner GCs which should have fallen to the center of Fornax and generated a nuclear cluster of ${\mathcal O}(10^6\, M_\odot)$. The absence of this nuclear cluster in Fornax may also have something to say about DM properties, though this is a different story.

We are fully aware that the dynamics of GCs is  more complex than what we presented in  \cite{Bar:2021jff}. Still, we believe that our simple analytical model captures some interesting (even intriguing) part of the dynamics, that we hope to study in the future with more realistic methods. 

\section{Conclusions and outlook}

In this presentation, we revisited the  timing `problem' of the GCs in the Fornax dSph, and use it to learn about new properties of DM. The latter are manifested in either a modification of the microscopic origin of dynamical friction (DF) or the halo morphologies, also influencing DF. Some of the 6 GCs of Fornax are placed too close to the center of this galaxy, which may pose a tuning problem in terms of typical time-scales to plunge to the center of this object. Indeed, the last part of the work described in  \cite{Bar:2021jff} (and briefly reviewed in Sec.~\ref{sec:CDF}) was devoted to quantifying this degree of tuning for the standard CDM paradigm yielding NFW profiles. The observations presented in Table~\ref{aba:tbl1} can be accounted for as a moderate fluctuation   with a Poisson probability of about  around $25\%$. 

Still, it is quite interesting that the calculation of DF in different DM models allows us to reduce this tension and find better accuracy with data. This is why a large part of this short presentation has consisted in a succinct explanation of how to derived DF from a Fokker-Planck formalism (which allows the introduction of fermionic and baryonic effects in the collision term).  As summarized in Fig.~\ref{fig:times},  the presence of a small core due to baryonic feedback may slightly alleviate this tension, but not at an interesting level. Once one considers other DM models (as DDM, ULDM and SIDM) a large core (of kpc size) may be formed, which predicts enough reduction of  dynamical friction to better reproduce the observed positions of the GCs in Fornax. Still, in these cases, the GC distribution depends strongly on initial conditions. From these models, it seems that SIDM with ${\sigma / m_{\rm DM}}\approx {1 \mathrm{~cm}^{2} / \mathrm{gr}}$ seems favoured, since it does not contradict any other observation, but more data is required to confirm this claim. 

The way forward is clear: applying our methods to more extensive data, aiming at enough statistical significance to generate robust conclusions about  DM properties. The hint about SIDM we just discussed makes this project particularly exciting, since future data from GCs may finally start closing up on the fundamental nature of DM.

\section*{Acknowledgments} It is a pleasure to thank N. Bar, K. Blum and H. Kim for all the hard work they devoted to this work.  IFAE is partially funded by the CERCA program of the Generalitat de Catalunya.  DB is supported by a `Ayuda Beatriz Galindo Senior' from the Spanish `Ministerio de Universidades', grant BG20/00228. 
The research leading to these results has received funding from the Spanish Ministry of Science and Innovation (PID2020-115845GB-I00/AEI/10.13039/501100011033).

\bibliographystyle{ws-procs961x669}
\bibliography{ref}

\end{document}